\pdfoutput=1
\documentclass[conference]{IEEEtran}
\IEEEoverridecommandlockouts

\usepackage{cite}
\usepackage{amsmath,amssymb,amsfonts}
\usepackage{algorithmic}
\usepackage{graphicx}
\usepackage{textcomp}
\usepackage{xcolor}
\def\BibTeX{{\rm B\kern-.05em{\sc i\kern-.025em b}\kern-.08em
    T\kern-.1667em\lower.7ex\hbox{E}\kern-.125emX}}
\usepackage{tabularx}
\usepackage{multirow}
\usepackage{footnote}
\usepackage{placeins}
\usepackage{hyperref}
\usepackage{amssymb}
\usepackage{amsmath}
\usepackage{amsfonts}
\hypersetup{
  colorlinks = true,
  linkcolor  = blue,
  urlcolor   = blue,
  citecolor  = black,
}
\usepackage{url}
\usepackage{cleveref}
\usepackage{thmbox}
\usepackage{mdframed}

\newcommand*{\fullref}[1]{\hyperref[{#1}]{\cref*{#1} \nameref*{#1}}}
\newcommand*{\Fullref}[1]{\hyperref[{#1}]{\Cref*{#1} \nameref*{#1}}}
\newcommand*{\secref}[1]{\hyperref[{#1}]{\autoref*{#1}}}
\newcommand*{\Secref}[1]{\hyperref[{#1}]{\Cref*{#1}}}

\let\oldFootnote\footnote
\newcommand\nextToken\relax

\renewcommand\footnote[1]{%
    \oldFootnote{#1}\futurelet\nextToken\isFootnote}

\newcommand\isFootnote{%
    \ifx\footnote\nextToken\textsuperscript{,}\fi}

\usepackage{booktabs}
\usepackage{xcolor}
\def\BibTeX{{\rm B\kern-.05em{\sc i\kern-.025em b}\kern-.08em
    T\kern-.1667em\lower.7ex\hbox{E}\kern-.125emX}}

\newcounter{openboxwithtitle}[section]

\begin{document}

\title{Fostering Knowledge Infrastructures in Science Communication and Aerospace Engineering 
\thanks{This work was funded by the DFG SE2A Excellence Cluster.}
}

\author{
    \IEEEauthorblockA{Tim Wittenborg\\L3S Research Center, Leibniz University Hannover, Hannover, Germany
    \\tim.wittenborg@l3s.uni-hannover.de}
}

\maketitle

\begin{abstract}
Knowledge infrastructures are defined as robust networks of people, artifacts, and institutions that generate, share and maintain specific knowledge.
Yet, many domains are fragmented and far from robustly networked, such as science communication or aerospace engineering.
While FAIR (\textbf{F}indable, \textbf{A}ccessible, \textbf{I}nteroperable, \textbf{R}eusable) data management tools exist, their adoption in these domains is limited.
Several challenges inhibit this adoption, from complex heterogeneous data formats to lack of structured support to outright incentives against collaboration or legal barriers.
This doctoral work outlines how to foster underdeveloped knowledge infrastructures with the use-cases of science communication and aerospace engineering.
By analyzing these problems and identifying available solutions, tool-supported workflows towards collaborative infrastructure can be implemented and evaluated.
These include human-in-the-loop artificial intelligence (AI)-supported workflows for information extraction and processing, wiki- and knowledge-graph-based digital libraries, and stakeholder-requirement-driven interfaces.
While these developed tools for workflow automation and knowledge representation show promise, significant challenges remain.
Future work will have to go beyond technical problem-solving and address the societal and legal barriers to unlock the particular domains.
Beyond that, advocates of emerging knowledge infrastructures in any domain are welcome to apply the findings of this work to foster the networking of available knowledge.
\end{abstract}

\begin{IEEEkeywords}
knowledge infrastructure, science communication, aerospace engineering, digital library, audiovisual content, wiki, media annotation, workflow automation
\end{IEEEkeywords}

\section{Research motivation}
Research in both science communication and aerospace engineering faces a common problem: valuable knowledge is produced continuously, but remains fragmented across formats, repositories, and organizational boundaries.
This fragmentation reduces all aspects of the FAIR Data principles: \textbf{F}indability, \textbf{A}ccessibility, \textbf{I}nteroperability and \textbf{R}eusability~\cite{wilkinson_fair_2016}.
Data-reliant processes become needlessly resource intensive, from simple video search to complex multidisciplinary design optimization.
At the same time, the state of the art steadily advances, in both knowledge-representation technologies, particularly knowledge graphs, and Workflow automation, especially human-in-the-loop artificial intelligence (AI) integration.
This thesis aims to foster emerging knowledge infrastructures by focussing on solutions that balance automation with human validation and incentivize collaborative curation practices in sustainable digital libraries. 
The driving research questions (RQ) are:
\begin{enumerate}
    \item [\textbf{RQ1:}] Which preferably domain-independent artifacts, particularly digital libraries and tools for workflow automation, are applicable to foster knowledge infrastructure?
    \item [\textbf{RQ2:}] How well can a modular tool-supported workflow framework meet the requirements for extracting, aligning and representing respective knowledge?
    \item [\textbf{RQ3:}] What domain specific requirements arise when applying knowledge infrastructure tools in science communication or aerospace engineering?
    \item [\textbf{RQ4:}] How can knowledge infrastructure requirements specific to fragmented domains such as science communication and aerospace engineering be met by digital libraries?
\end{enumerate}

\section{Background and Related Work\label{sec:related}}
Knowledge Infrastructures (KI) are defined by Edwards~\cite{10.5555/1805940} as ``\textit{robust networks of people, artifacts, and institutions that generate, share, and maintain specific knowledge about the human and natural worlds}''.
Notable examples include the Intergovernmental Panel on Climate Change (IPCC) and the Wikipedia Community, with its individual contributors, formal and informal organizations, and (sub-)projects, digital libraries, tools, etc.
These networks are characterized by enduring beyond any individual project time~\cite{karasti_infrastructure_2010}.
In domains such as science communication and aerospace engineering, however, such robust networks do not exist, and efforts are fragmented, particularly when considering the FAIR Data principles~\cite{wilkinson_fair_2016}. 
However, uniting these endeavors requires not only technical components and standards, but also conventions for provenance, governance, and community adoption towards participation that ensure long-term quality and reuse.
As such, this section highlights common artifacts, most notably \nameref{sec:dl}, and specifies their adoption in two domain applications: \nameref{sec:scicom} and \nameref{sec:aerospace}.

\subsection{Digital Libraries\label{sec:dl}}
Definitions of Digital Libraries (DLs) vary, from ``\textit{a set of electronic resources and associated technical capabilities for creating, searching and using information}'' to including the staff and community working on it~\cite{borgman_what_1999}.
Focussing on the artifact component of it, modern DL increasingly rely on structured, machine-actionable data, such as knowledge graphs.
Notable examples of such graph-based digital libraries are Wikidata, presumably the world's largest collaboratively curated knowledge graph~\cite{vrandecic_wikidata_2014}.
DLs with such a vast scope quickly face scaling issues and meet their limit before meeting the needs of their communities\footnote{\url{https://www.wikidata.org/wiki/Wikidata:WikiProject_Limits_of_Wikidata}}, which is why efforts have transitioned from monolithic to a federated ecosystem.
This ecosystem surrounding Wikidata, encompassing more than a thousand specialized wikibase instances, is called Wikiverse, which is again part of the Linked Open Data (LOD) Environment.
In this LOD, another knowledge graph warrants note, the Open Research Knowledge Graph (ORKG)~\cite{auer_improving_2020}, which focuses on semantic representation of research contributions.
These artifacts are of universal interest to the scientific method and potentially every knowledge infrastructure.
Since Wikidata and the ORKG are advancing the state-of-the-art of DL every day, this work focusses on aiding underdeveloped infrastructures to catch up to this evolving potential.

\subsection{Science Communication\label{sec:scicom}}
Burns et al.~\cite{burns_science_2003} define Science Communication (SciCom) as ``\textit{the use of appropriate skills, media, activities, and dialogue to produce one or more of the following personal responses to science (the AEIOU vowel analogy): Awareness, Enjoyment, Interest, Opinion-forming, and Understanding}''.
It aims to improve public understanding, acceptance, trust, and support while providing a venue to gather broad feedback, local knowledge, and civic needs regarding valuable research aims and applications~\cite{kappel_why_2019}.
The SciCom KI is then the knowledge infrastructure dedicated to science communication, its people, institutions and artifacts, such as videos and podcasts.
MacKenzie~\cite{mackenzie_science_2019} has curated 952 science podcasts, \href{https://wissenschaftspodcasts.de/podcasts/}{wissenschaftspodcasts.de} 386; Kikkawa et al.~\cite{kikkawa_enhancing_2024,kikkawa_ya_2024} 230,000 videos, the world lecture project\footnote{\url{https://world-lecture-project.org/}} 59,634, the TIB-AV Portal~\cite{marin_arraiza_tibav_2015} 50,690.
Such videos are meticulously researched, with hundreds of different qualities and properties being analyzed and annotated~\cite{navarrete_closer_2023}.
This secondary data is usually stored in isolated repositories, disconnected from the original media, unless it is captured within FAIR infrastructure -- for instance, through licensing media in the TIB AV-Portal~\cite{marin_arraiza_tibav_2015} or by meeting the notability criteria\footnote{\url{https://www.wikidata.org/wiki/Wikidata:Notability}} of Wikidata~\cite{vrandecic_wikidata_2014}.
Yet, capturing this knowledge is especially relevant when assessing information that coexists within the same communication channels, but varies in quality, purpose, scope, evaluation standards, accessibility, and revenue generation~\cite{fahnrich_exploring_2023,hagenhoff_neue_2007}.

\subsection{Aerospace Engineering\label{sec:aerospace}}
The aerospace domain relies on formalized engineering processes and Knowledge‑Based Engineering (KBE) to capture expert rules and automate recurring design, analysis, and documentation tasks~\cite{stjepandic_knowledge-based_2015,reddy_knowledge_2015,quintana2015transformingKBE}. As systems grow more multidisciplinary, reuse of heterogeneous knowledge from explicit rules and results to best practices and guidelines becomes critical for managing complexity and accelerating iterations~\cite{la_rocca_FE-models_2007,staack2013systemKBEdesign}.
Semantic technologies and ontologies are increasingly used to improve interoperability and enable automated reasoning and discovery across tools and teams~\cite{knoos_franzen_system_2023,dadzie_applying_2009}.
Central exchange formats are developed, such as CPACS or CMDOWS~\cite{van_gent_cmdows_2018}.
Yet, early domain studies have long since listed prerequisites and challenges for next‑generation aerospace knowledge bases: 
\textit{``It was established that improving the structure and representation of knowledge would benefit and increase the competency of design engineering activities''}~\cite{sanya_challenges_2011}.
They specifically mention that \textit{``engineers often spend large amounts of time searching for a solution to problems that may have been solved.''}, partially because \textit{``engineers are sometimes reluctant to share documents''}.
Concerns about intellectual property and governance further restrain sharing, state Procko and Ochoa~\cite{procko_leveraging_2022} in their literature review, citing  Harvey and Holdsworth~\cite{harvey_knowledge_2005}:
\textit{``In addition to these challenges, the aerospace and defense industries are perhaps the most governed of all sectors, being regulated by a plethora of agencies, both national and regional.''}

This thesis aims to summarize these findings and provide actionable solutions towards fostering knowledge infrastructure in this sharing-hazardous environment.
Modular solutions need to combine KBE requirements with expert oversight and preserve restrictive governance needs.
The goal is to foster collaborative, platform-agnostic, linked open data infrastructure in science communication and aerospace engineering.

\section{Research approach and methodology}
\begin{figure*}
    \centering
    \includegraphics[width=1.0\linewidth]{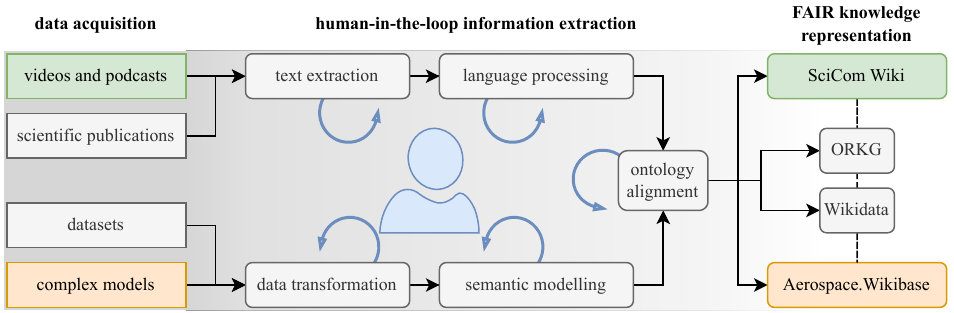}
    \caption{Overview of the workflow, processing data towards FAIR knowledge representation. Audiovisual content can be similarly processed as textual publications. KBE models and datasets in general equally benefit from semantic information extraction and knowledge representation as Linked (Open) Data.}
    \label{fig:overview}
\end{figure*}

This work uses a mixed-methods research approach, building on systematic literature reviews (SLRs), requirements elicitation, and action-research-driven software development.
The foundation is depicted in \secref{fig:overview}, the development of a data processing pipeline, that can be used to facilitate systematic literature reviews, but also incorporates modules that can be repurposed for other applications.
An aerospace SLR can display the pipeline's capability and inform the future development towards establishing an aerospace engineering KI.
The underlying pipeline, however, can be used to process audiovisual media, such as videos and podcasts, and enrich corpus analysis similarly to the SLR.
Once structured like this, further applications become feasible, such as computational fact-checking against similarly processed knowledge bases.

This approach arose dynamically, meeting emerging requirements and utilizing identified synergies.
Particularly, learning from related work shaped this development, from domain-specifics such as CMDOWS seeking to formalize workflows, to external inspirations such as federated learning aiming to produce cohesive improvement in similarly fragmented environments. 
For development purposes, action research with domain expert evaluation was used.
The evaluation was facilitated through surveys and interviews, which also complemented literature review findings.

\section{Preliminary Results}

\subsection{SWARM-SLR}
SWARM-SLR (Streamlined Workflow Automation for Machine-actionable Systematic Literature Reviews) is a modular, requirements-driven workflow that formalizes 65 SLR requirements and maps task-specialized tools, such as reference managers, natural language processing (NLP) pipelines and digital libraries, to the full SLR lifecycle from planning through reporting.
Structured as four Jupyter notebooks, it guides through the SLR with human-in-the-loop validation to enable interoperable and reusable semi-automated literature reviews.
The paper is published~\cite{wittenborg_swarm-slr_2024}.
\paragraph{ExtracTable}
ExtracTable refines the corpus creation and data extraction workflow of SWARM-SLR.
It is designed for downstream integration into knowledge graphs (KGs) by combining large language models (LLMs) with human-in-the-loop-defined schemas.
The paper is accepted~\cite{john_extractable_2026}.

\subsection{Knowledge Based Aerospace Engineering}
Using SWARM-SLR, we screened over 1.000 documents and performed a detailed analysis of 164 core articles to map current KBE practices in aerospace.
We find a solid technical base, especially CPACS and CMDOWS, but fragmented adoption.
Existing literature recommends prioritizing open exchange formats, collaborative data environments and commitments to overcome data silos. The reuse of data should be compatible with the restrictive governance needs of aerospace engineering.
The journal article is under review~\cite{wittenborg_knowledge-based_2025}.

\subsection{SciCom Wiki}
The SciCom Wiki is an open, Wikibase-based digital library for scientific videos and podcasts, designed from stakeholder requirements and prototype microservices to improve discovery, annotation, and FAIR metadata capture for non-textual media.
Early evaluations (surveys, interviews and prototype testing) indicate it substantially aids findability and curation while identifying legal, onboarding, and scalability tasks required for broader adoption.
The paper is accepted~\cite{wittenborg_scicom_2025}.

\paragraph{Computational Fact-Checking}
A computational fact-checking prototype pairs LLM-based statement extraction with symbolic verification against curated ground-truth knowledge graphs to produce interpretable scientific-accuracy scores for media content. Expert and user evaluations show the approach provides useful veracity indications but is currently limited by non-FAIR ground-truth representation and inefficient extraction.
The paper is accepted~\cite{wittenborg_computational_2025}.

\section{Challenges and Discussion}
This work aims to answer how tool-supported workflows towards FAIR data can foster knowledge infrastructures. Several findings and identified challenges warrant future work:

\textbf{RQ1} asks for artifacts to foster knowledge infrastructure, and plenty suitable candidates could be identified.
From large graph-based digital libraries to small dedicated instances, well established standards and emerging interoperable exchange formats and workflow automation tools, these artifacts form a comprehensive catalogue.
Community uptake, however, remains challenging throughout most of them.

\textbf{RQ2} sought to measure the capability of a modular tool-supported workflow framework, where SWARM-SLR shows that a notebook-driven framework with human-in-the-loop validation can satisfy many requirements regarding knowledge extraction, structuring, and representation. 
Some adoption and usability challenges may be answered with ExtracTable and SWARM-SLR AIssistant, yet the abstraction and AI-tradeoffs leave the current state with much room to grow before all requirements are sufficiently met.

\textbf{RQ3} seeks domain-specific KI requirements.
The domain of science communication has accessible audiovisual data that requires structure, straightforward, yet not trivial legal considerations, and a need to consolidate contributors.
Aerospace engineering has various structured approaches and schema, yet is held back by self-afflicted constraints. 
The core challenge is designing solutions that respect regulations while facilitating interoperability.

\textbf{RQ4} challenges how digital libraries can meet domain requirements.
The aerospace.wikibase is a first step towards a common foundation in its highly restricted domain, providing at least a common taxonomy and persistent identifiers.
The SciCom Wiki, however, can provide the same and go much further beyond, towards an openly accessible digital library, since data sharing is not an issue in the science communication domain.
Evidently, solutions built on federated LOD instances, particularly in the Wikibase ecosystem, can meet many requirements.
Plenty of available artifacts are readily available, and modular worfklow frameworks exist and can be customized.
These findings are preliminary and reflect ongoing work, and future iterations and evaluations may reveal more than the current coverage.

\section*{Acknowledgment}
\paragraph*{Use of AI tools declaration}
During the preparation of this work, the author(s) used \textbf{GPT-5 mini (GitHub Copilot)}, \textbf{DeepL}, \textbf{Grammarly (Browser Plugin)}, \textbf{LanguageTool (Browser Plugin)} in order to: \textbf{translate text}, \textbf{grammar and spelling check}, \textbf{paraphrase and reword}, according to the CEUR GenAI Usage Taxonomy\footnote{\url{https://ceur-ws.org/GenAI/Taxonomy.html}}.
After using this tool/service, the authors reviewed and edited the content as needed and take full responsibility for the publication’s content.

\bibliographystyle{ieeetr}
\bibliography{main}

@article{wilkinson_fair_2016,
	title = {The {FAIR} {Guiding} {Principles} for scientific data management and stewardship},
	volume = {3},
	copyright = {2016 The Author(s)},
	issn = {2052-4463},
	doi = {10.1038/sdata.2016.18},
	language = {en},
	number = {1},
	urldate = {2025-01-18},
	journal = {Scientific Data},
	author = {Wilkinson, Mark D. and Dumontier, Michel and Aalbersberg, IJsbrand Jan and Appleton, Gabrielle and Axton, Myles and Baak, Arie and other},
	month = mar,
	year = {2016},
	note = {Publisher: Nature Publishing Group},
}

@book{10.5555/1805940,
    author = {Edwards, Paul N.},
    title = {A Vast Machine: Computer Models, Climate Data, and the Politics of Global Warming},
    year = {2010},
    isbn = {0262013924},
    publisher = {The MIT Press}
}

@article{karasti_infrastructure_2010,
	title = {Infrastructure {Time}: {Long}-term {Matters} in {Collaborative} {Development}},
	volume = {19},
	issn = {1573-7551},
	shorttitle = {Infrastructure {Time}},
	doi = {10.1007/s10606-010-9113-z},
	language = {en},
	number = {3},
	urldate = {2025-03-10},
	journal = {Computer Supported Cooperative Work (CSCW)},
	author = {Karasti, Helena and Baker, Karen S. and Millerand, Florence},
	month = aug,
	year = {2010},
	pages = {377--415},
}

@article{borgman_what_1999,
	title = {What are digital libraries? {Competing} visions},
	volume = {35},
	issn = {0306-4573},
	shorttitle = {What are digital libraries?},
	url = {https://escholarship.org/uc/item/7c55m1xf},
	doi = {10.1016/S0306-4573(98)00059-4},
	language = {en},
	number = {3},
	urldate = {2025-11-05},
	journal = {Information Processing \& Management},
	author = {Borgman, Christine L.},
	year = {1999},
	pages = {227--243},
}

@article{vrandecic_wikidata_2014,
	title = {Wikidata: a free collaborative knowledgebase},
	volume = {57},
	issn = {0001-0782},
	shorttitle = {Wikidata},
	doi = {10.1145/2629489},
	number = {10},
	urldate = {2025-03-10},
	journal = {Commun. ACM},
	author = {Vrandečić, Denny and Krötzsch, Markus},
	month = sep,
	year = {2014},
	pages = {78--85},
}

@article{auer_improving_2020,
	title = {Improving {Access} to {Scientific} {Literature} with {Knowledge} {Graphs}},
	volume = {44},
	copyright = {De Gruyter expressly reserves the right to use all content for commercial text and data mining within the meaning of Section 44b of the German Copyright Act.},
	issn = {1865-7648},
	url = {https://www.degruyterbrill.com/document/doi/10.1515/bfp-2020-2042/html},
	doi = {10.1515/bfp-2020-2042},
	language = {en},
	number = {3},
	urldate = {2025-09-24},
	journal = {Bibliothek Forschung und Praxis},
	author = {Auer, Sören and Oelen, Allard and Haris, Muhammad and Stocker, Markus and D’Souza, Jennifer and Farfar, Kheir Eddine and Vogt, Lars and Prinz, Manuel and Wiens, Vitalis and Jaradeh, Mohamad Yaser},
	month = dec,
	year = {2020},
	note = {Publisher: De Gruyter},
	pages = {516--529},
}

@article{burns_science_2003,
	title = {Science {Communication}: {A} {Contemporary} {Definition}},
	volume = {12},
	issn = {0963-6625},
	shorttitle = {Science {Communication}},
	doi = {10.1177/09636625030122004},
	language = {en},
	number = {2},
	urldate = {2025-03-07},
	journal = {Public Understanding of Science},
	author = {Burns, T. W. and O'Connor, D. J. and Stocklmayer, S. M.},
	month = apr,
	year = {2003},
	note = {Publisher: SAGE Publications Ltd},
	pages = {183--202},
}

@article{kappel_why_2019,
	title = {Why {Science} {Communication}, and {Does} {It} {Work}? {A} {Taxonomy} of {Science} {Communication} {Aims} and a {Survey} of the {Empirical} {Evidence}},
	volume = {4},
	issn = {2297-900X},
	shorttitle = {Why {Science} {Communication}, and {Does} {It} {Work}?},
	doi = {10.3389/fcomm.2019.00055},
	language = {English},
	urldate = {2025-03-07},
	journal = {Frontiers in Communication},
	author = {Kappel, Klemens and Holmen, Sebastian Jon},
	month = oct,
	year = {2019},
	note = {Publisher: Frontiers},
}

@article{mackenzie_science_2019,
	title = {Science podcasts: analysis of global production and output from 2004 to 2018},
	copyright = {© 2019 The Authors.},
	shorttitle = {Science podcasts},
	doi = {10.1098/rsos.180932},
	language = {EN},
	urldate = {2025-03-26},
	journal = {Royal Society Open Science},
	author = {MacKenzie, Lewis E.},
	month = jan,
	year = {2019},
	note = {Publisher: The Royal Society Publishing},
}

@inproceedings{kikkawa_enhancing_2024,
	address = {Cham},
	title = {Enhancing {Identification} of {Scholarly} {Reference} on {YouTube}},
	isbn = {978-3-031-72437-4},
	doi = {10.1007/978-3-031-72437-4_19},
	language = {en},
	booktitle = {Linking {Theory} and {Practice} of {Digital} {Libraries}},
	publisher = {Springer Nature Switzerland},
	author = {Kikkawa, Jiro and Takaku, Masao and Yoshikane, Fuyuki},
	editor = {Antonacopoulos, Apostolos and Hinze, Annika and Piwowarski, Benjamin and Coustaty, Mickaël and Di Nunzio, Giorgio Maria and Gelati, Francesco and Vanderschantz, Nicholas},
	year = {2024},
	pages = {326--341},
}

@misc{kikkawa_ya_2024,
	title = {{YA} {Domain} {Dataset}: {Dataset} of scholarly bibliographic references on {YouTube} videos},
	shorttitle = {{YA} {Domain} {Dataset}},
	url = {https://zenodo.org/records/12801387},
	urldate = {2025-08-22},
	publisher = {Zenodo},
	author = {Kikkawa, Jiro and Takaku, Masao and Yoshikane, Fuyuki},
	month = jul,
	year = {2024},
}

@inproceedings{marin_arraiza_tibav_2015,
	address = {New York, NY, USA},
	series = {{WWW} '15 {Companion}},
	title = {The {TIB}{\textbar}{AV} {Portal} as a {Future} {Linked} {Media} {Ecosystem}},
	isbn = {978-1-4503-3473-0},
	doi = {10.1145/2740908.2742912},
	urldate = {2025-03-10},
	booktitle = {Proceedings of the 24th {International} {Conference} on {World} {Wide} {Web}},
	publisher = {Association for Computing Machinery},
	author = {Marín Arraiza, Paloma and Strobel, Sven},
	month = may,
	year = {2015},
	pages = {733--734},
}

@misc{navarrete_closer_2023,
	title = {A {Closer} {Look} into {Recent} {Video}-based {Learning} {Research}: {A} {Comprehensive} {Review} of {Video} {Characteristics}, {Tools}, {Technologies}, and {Learning} {Effectiveness}},
	shorttitle = {A {Closer} {Look} into {Recent} {Video}-based {Learning} {Research}},
	doi = {10.48550/arXiv.2301.13617},
	urldate = {2025-03-07},
	publisher = {arXiv},
	author = {Navarrete, Evelyn and Nehring, Andreas and Schanze, Sascha and Ewerth, Ralph and Hoppe, Anett},
	month = aug,
	year = {2023},
	note = {arXiv:2301.13617 [cs]},
}

@article{fahnrich_exploring_2023,
	title = {Exploring ‘quality’ in science communication online: {Expert} thoughts on how to assess and promote science communication quality in digital media contexts},
	volume = {32},
	issn = {0963-6625},
	shorttitle = {Exploring ‘quality’ in science communication online},
	doi = {10.1177/09636625221148054},
	language = {en},
	number = {5},
	urldate = {2025-03-07},
	journal = {Public Understanding of Science},
	author = {Fähnrich, Birte and Weitkamp, Emma and Kupper, J. Frank},
	month = jul,
	year = {2023},
	note = {Publisher: SAGE Publications Ltd},
	pages = {605--621},
}

@book{hagenhoff_neue_2007,
	title = {Neue {Formen} der {Wissenschaftskommunikation}: eine {Fallstudienuntersuchung}},
	isbn = {978-3-938616-75-8},
	shorttitle = {Neue {Formen} der {Wissenschaftskommunikation}},
	language = {German},
	urldate = {2025-01-25},
	publisher = {Universitätsverlag Göttingen},
	author = {Hagenhoff, Svenja and Seidenfaden, Lutz and Ortelbach, Björn and Schumann, Matthias},
	year = {2007},
	doi = {10.17875/gup2007-208},
}

@incollection{stjepandic_knowledge-based_2015,
	address = {Cham},
	title = {Knowledge-{Based} {Engineering}},
	isbn = {978-3-319-13776-6},
	url = {https://doi.org/10.1007/978-3-319-13776-6_10},
	language = {en},
	urldate = {2025-11-21},
	booktitle = {Concurrent {Engineering} in the 21st {Century}: {Foundations}, {Developments} and {Challenges}},
	publisher = {Springer International Publishing},
	author = {Stjepandić, Josip and Verhagen, Wim J. C. and Liese, Harald and Bermell-Garcia, Pablo},
	year = {2015},
	doi = {10.1007/978-3-319-13776-6_10},
	pages = {255--286},
}

@article{reddy_knowledge_2015,
	title = {Knowledge {Based} {Engineering}: {Notion}, {Approaches} and {Future} {Trends}},
	volume = {5},
	url = {https://www.semanticscholar.org/paper/01a20390b48152ec02dc0bf29b109aded7e0d9a5},
	journal = {American Journal of Intelligent Systems},
	author = {Reddy, E. J. and Sridhar, C. and Rangadu, V.},
	year = {2015},
	pages = {1--17},
}

@article{quintana2015transformingKBE,
    title = {Transforming expertise into Knowledge-Based Engineering tools: A survey of knowledge sourcing in the context of engineering design},
    author = {Quintana-Amate, S. and  Bermell-Garcia, P. and Tiwari, A.},
    journal = {Knowledge-Based Systems},
    volume = {84},
    pages = {89-97},
    year = {2015},
    issn = {0950-7051},
    doi = {https://doi.org/10.1016/j.knosys.2015.04.002},
    url = {https://www.sciencedirect.com/science/article/pii/S0950705115001409},
    comment={IS: used for reference listing KBE methods and old(!) trends}
}

@Conference{la_rocca_FE-models_2007,
  author =    {La Rocca, Gianfranco and van Tooren, Michel J.L.},
  title =     {A Knowledge Based Engineering Approach to Support Automatic Generation of {FE} Models in Aircraft Design},
  booktitle = {45th AIAA Aerospace Sciences Meeting and Exhibit, Aerospace Sciences Meetings},
  year =      {2007},
  address =   {Reno, Nevada},
  month =     {8-11 January}
}

@InProceedings{staack2013systemKBEdesign,
    author =       {Staack, Ingo and Krus, Petter},
    title =        {Integration of On-Board Power Systems Simulation in Conceptual Aircraft Design},
    booktitle =    {Proceedings of the 4th CEAS European Air \& Space Conference},
    year =         {2013},
    pages =        {709},
    organization = {Council of the European Aerospace Societies (CEAS)},
    publisher =    {Linköping University Electronic Press, 2013},
    url =          {https://login.e.bibl.liu.se/login?url=https://search.ebscohost.com/login.aspx?direct=true&db=edsswe&AN=edsswe.oai.DiVA.org.liu.103197&site=eds-live&scope=site}
}

@phdthesis{knoos_franzen_system_2023,
	type = {{PhD} {Thesis}},
	title = {A {System} of {Systems} {View} in {Early} {Product} {Development} : {An} {Ontology}-{Based} {Approach}},
	school = {Linköping UniversityLinköping University, Fluid and Mechatronic Systems, Faculty of Science \& Engineering},
	author = {Knöös Franzén, Ludvig},
	year = {2023},
	doi = {10.3384/9789180751667},
}

@article{dadzie_applying_2009,
    title = {Applying semantic web technologies to knowledge sharing in aerospace engineering},
    volume = {20},
    issn = {0956-5515, 1572-8145},
    url = {http://link.springer.com/10.1007/s10845-008-0141-1},
    doi = {10.1007/s10845-008-0141-1},
    language = {en},
    number = {5},
    urldate = {2024-02-27},
    journal = {Journal of Intelligent Manufacturing},
    author = {Dadzie, A.-S. and Bhagdev, R. and Chakravarthy, A. and Chapman, S. and Iria, J. and Lanfranchi, V. and Magalhães, J. and Petrelli, D. and Ciravegna, F.},
    year = {2009},
    pages = {611--623},
}

@article{van_gent_cmdows_2018,
	title = {{CMDOWS}: a proposed new standard to store and exchange {MDO} systems},
	volume = {9},
	issn = {1869-5590},
	shorttitle = {{CMDOWS}},
	url = {https://doi.org/10.1007/s13272-018-0307-2},
	doi = {10.1007/s13272-018-0307-2},
	language = {en},
	number = {4},
	urldate = {2025-11-05},
	journal = {CEAS Aeronautical Journal},
	author = {van Gent, Imco and La Rocca, Gianfranco and Hoogreef, Maurice F. M.},
	month = dec,
	year = {2018},
	pages = {607--627},
}

@inproceedings{sanya_challenges_2011,
	title = {Challenges in semantic knowledge management for aerospace design engineering},
	url = {https://www.designsociety.org/publication/30631/challenges_in_semantic_knowledge_management_for_aerospace_design_engineering},
	urldate = {2024-02-27},
	booktitle = {{DS} 68-6: {Proceedings} of the 18th {International} {Conference} on {Engineering} {Design} ({ICED} 11), {Impacting} {Society} through {Engineering} {Design}, {Vol}. 6: {Design} {Information} and {Knowledge}, {Lyngby}/{Copenhagen}, {Denmark}, 15.-19.08. 2011},
	author = {Sanya, Isaac and Shehab, Essam and Lowe, Dave},
	year = {2011},
	pages = {241--248},
}

@article{procko_leveraging_2022,
	title = {Leveraging {Linked} {Data} for {Knowledge} {Management}: {A} {Proposal} for the {Aerospace} {Industry}},
	shorttitle = {Leveraging {Linked} {Data} for {Knowledge} {Management}},
	url = {https://papers.ssrn.com/sol3/papers.cfm?abstract_id=4703237},
	urldate = {2024-02-27},
	journal = {Available at SSRN 4703237},
	author = {Procko, Tyler and Ochoa, Omar},
	year = {2022},
}

@inproceedings{harvey_knowledge_2005,
	title = {Knowledge management in the aerospace industry},
	url = {https://ieeexplore.ieee.org/document/1494182},
	doi = {10.1109/IPCC.2005.1494182},
	urldate = {2025-01-26},
	booktitle = {{IPCC} 2005. {Proceedings}. {International} {Professional} {Communication} {Conference}, 2005.},
	author = {Harvey, D.J. and Holdsworth, R.},
	month = jul,
	year = {2005},
	note = {ISSN: 2158-1002},
	pages = {237--243},
}

@inproceedings{wittenborg_swarm-slr_2024,
	address = {Cham},
	title = {{SWARM}-{SLR} - {Streamlined} {Workflow} {Automation} for {Machine}-{Actionable} {Systematic} {Literature} {Reviews}},
	isbn = {978-3-031-72437-4},
	doi = {10.1007/978-3-031-72437-4_2},
	language = {en},
	booktitle = {Linking {Theory} and {Practice} of {Digital} {Libraries}},
	publisher = {Springer Nature Switzerland},
	author = {Wittenborg, Tim and Karras, Oliver and Auer, Sören},
	editor = {Antonacopoulos, Apostolos and Hinze, Annika and Piwowarski, Benjamin and Coustaty, Mickaël and Di Nunzio, Giorgio Maria and Gelati, Francesco and Vanderschantz, Nicholas},
	year = {2024},
	pages = {20--40},
}

@inproceedings{john_extractable_2026,
	address = {Cham},
	title = {{ExtracTable}: {Human}-in-the-{Loop} {Transformation} of {Scientific} {Corpora} into {Structured} {Knowledge}},
	isbn = {978-3-032-05409-8},
	shorttitle = {{ExtracTable}},
	doi = {10.1007/978-3-032-05409-8_27},
	language = {en},
	booktitle = {Linking {Theory} and {Practice} of {Digital} {Libraries}},
	publisher = {Springer Nature Switzerland},
	author = {John, Lena and Ghanmi, Ahmed Malek and Wittenborg, Tim and Auer, Sören and Karras, Oliver},
	editor = {Balke, Wolf-Tilo and Golub, Koraljka and Manolopoulos, Yannis and Stefanidis, Kostas and Zhang, Zheying},
	year = {2026},
	pages = {470--487},
}

@misc{wittenborg_knowledge-based_2025,
	title = {Knowledge-{Based} {Aerospace} {Engineering} -- {A} {Systematic} {Literature} {Review}},
	url = {http://arxiv.org/abs/2505.10142},
	doi = {10.48550/arXiv.2505.10142},
	urldate = {2025-11-05},
	publisher = {arXiv},
	author = {Wittenborg, Tim and Baimuratov, Ildar and Franzén, Ludvig Knöös and Staack, Ingo and Römer, Ulrich and Auer, Sören},
	month = may,
	year = {2025},
	note = {arXiv:2505.10142 [cs]},
}

@misc{wittenborg_scicom_2025,
	title = {{SciCom} {Wiki}: {A} {Digital} {Library} to {Support} the {Science} {Communication} {Knowledge} {Infrastructure} for {Videos} and {Podcasts}},
	shorttitle = {{SciCom} {Wiki}},
	url = {http://arxiv.org/abs/2511.09248},
	doi = {10.48550/arXiv.2511.09248},
	urldate = {2025-11-19},
	publisher = {arXiv},
	author = {Wittenborg, Tim and Stehr, Niklas and Karras, Oliver and Auer, Sören},
	month = nov,
	year = {2025},
	note = {arXiv:2511.09248 [cs]},
}

@inproceedings{wittenborg_computational_2025,
	title = {{Computational} {Fact}-{Checking} of {Online} {Discourse}: {Scoring} {Scientific} {Accuracy} in {Climate} {Change} {Related} {News} {Articles}},
	doi = {DOI 10.1109/ICKG66886.2025.00055},
	booktitle = {2025 {IEEE} {International} {Conference} on {Knowledge} {Graph} ({ICKG})},
	author = {Wittenborg, Tim and Tremel, Constantin Sebastian and Karras, Oliver and Auer, Sören},
	year = {2025},
	pages = {371--378},
}

\end{document}